\documentclass[12pt]{iopart}

\usepackage{graphicx}

\usepackage{psfrag,pstricks}
\begin{document}

\title[]{Optical Single Photons on Demand Teleported from Microwave Cavities}
\author{ Sh. Barzanjeh$^{1}$, G. J. Milburn$^{2}$, D. Vitali$^{1}$ and P. Tombesi$^{1}$}


\address{$^{1}$ Sezione di Fisica, Scuola di Scienze e Tecnologie, Universit\`{a} di Camerino, \textit{I-62032} Camerino (MC), Italy\\
$^{2}$Department of Physics, School of Physical Sciences, The University of Queensland, Saint Lucia, QLD 4072, Australia}
\ead{shabirbarzanjeh@gmail.com}
\begin{abstract}

We propose a scheme for entangling the optical and microwave output modes of the respective cavities by using a micro mechanical resonator. The micro-mechanical resonator on one side
is capacitively coupled to the microwave cavity and on the other side is coupled to a high finesses optical cavity. We then show how this continuous variable entanglement can be profitably used to teleport the non-Gaussian number state $|1\rangle$ and the superposition $(|0\rangle+|1\rangle)/\sqrt 2$ from the microwave cavity output mode onto an output of optical cavity mode  with fidelity much larger than the no-cloning limit.
\end{abstract}


\maketitle

\section{Introduction}\label{Introduction}
The on demand generation of optical single photons on a chip is one of the most challenging and required results for the successful implementation of quantum information devices. Many proposal for the production of single optical photons have been described and realized in the last decades. Early experiments demonstrated photon generation
from single ions \cite{walther}, atoms \cite{rempe1, rempe2, darquie}, and  molecules \cite{lounis}. The challenges involved in overcoming the practical difficulties in isolating single particles make their use as single photon sources very demanding. The first demonstration of stable, triggered, room temperature single-photon source was made using a nickel nitrogen defect \cite{grangier}. Another alternative is to use  quantum dots \cite{intallara}, though radiation in all directions makes  efficient collection difficult.
It is also possible to use twin photons produced in parametric down-conversion
to generate a 'heralded' source of single photons, 'heralded'
meaning that the single pohoton state is conditional on the detection of the other photon of the pair. The production of
1550 nm wavelength photons in this way was reported \cite{gisin, castelletto, sasaki}. Perhaps one of the largest disadvantages of the single-photon
sources described so far is that the emission is random. It
is not possible to tell if a particular excitation pulse has
generated a single-photon emission until that single photon
is detected.

A true resource for quantum information on a chip is the mapping of qubit states onto microwave photon states. These photons are
generated on-demand with a high repetition rate, high efficiency,
and good spectral purity \cite{martinis, houck}. The recent production
of single microwave photons, and superpositions of photon
states into an LC resonator from a superconducting flux qubit \cite{martinis, kosuke}, and the creation of a microwave photon counter \cite{mcdermott} are important steps towards on-chip quantum optics experiments.

In this work we will show instead that the single microwave photon generated on-demand in superconducting cavities can be teleported with high fidelity into a single photon at optical wavelength exploiting the entanglement between the output radiation mediated by a mechanical resonator. 

Entanglement is the property possessed by a multipartite
quantum system when it is in a state that cannot be factorized
into a product of states or a mixture of such products. In an
entangled state the various parties share nonclassical and possibly
non-local correlations, that are at the heart of counter intuitive  quantum phenomenon. Recently, wide range of experimental and theoretical schemes have been proposed to observe entanglement also in macroscopic
objects \cite{blen}, for example, the proposed entanglement of two mirrors of a ring cavity  by using the radiation
pressure of the cavity mode \cite{mancini1}. Subsequently, different schemes have been proposed to entangle of nano- and micro mechanical resonators with macroscopic and microscopic systems. such as the entanglement of a nanomechanical
resonator with a Cooper pair box \cite{armour}, or an
optical mode \cite{vitali1}, for entangling two charge qubits \cite{zou} or two
Josephson junctions \cite{cleland} via nanomechanical resonators, and
for entangling two nanomechanical resonators via trapped
ions \cite{tian1} or Cooper pair boxes \cite{tian2, ring}. 

Owing to the recent improvements in nanofabrication techniques a new scheme for entangling a nano-mechanical resonator
with the microwave field of a superconducting coplanar waveguide field, without the mediation of a Cooper pair box, was proposed \cite{vitali2}.
In particular, this continuous-variable (CV) entanglement can be used to teleport an unknown quantum state. Quantum teleportation  \cite{benn} is the transfer of an unknown quantum state from a sender (Alice) to a receiver (Bob) by
means of the entanglement shared by the two parties and appropriate classical communication. The teleportation is perfect and Bob recovers an exact
copy of the state teleported to him by Alice only if the
quantum channel is an ideal maximally entangled state. If
we deal with qubits represented by polarization states
of photons, then we can employ pair of polarization entangled photons generated by means of spontaneous
parametric down-conversion, wherein the entanglement is
almost perfect \cite{bo,bosc}. However, in the case of continuous
quantum variables \cite{vai,bra}, an ideal channel is an unphysical infinitely squeezed state. In quantum optics, by
considering the finite quantum correlations between the
quadratures in a two-mode squeezed state, Braunstein and
Kimble \cite{bra} proposed a realistic protocol employing a beam
splitter and homodyne measurements, which approaches perfect teleportation in the limit of infinite degree of squeezing. This teleportation was first realized in Ref. \cite{furusawa1} using a Gaussian coherent state then was successfully extended to a non-Gaussian state in Ref. \cite{furusawa2}

In Ref.\cite{bar} we proposed a specific optomechanical system for the teleportation of Schr\"{o}dingers-cat states. In the present paper we show how a CV quantum teleportation protocol can be implemented in the same optomechanical system for realizing the teleportation of a single photon state and even of a coherent superposition of number states from microwave to optical frequencies.  Combining this scheme with the demonstrated ability to generate  on demand single microwave photons, we could realise a deterministic source of single optical photons. 

We consider a hybrid, strongly quantum-correlated system formed by a microwave
cavity (MC) coupled to a high-finesse optical cavity (OC) via a vibrating micro cantilever. The microwave cavity mode is indirectly coupled to an optical cavity mode via the common
interaction with the vibrating micro mechanical resonator \cite{barzanjeh}. We
show that with the current scheme, it is possible
to generate a reversible stationary CV
entanglement between the output fields of optical and microwave resonators, which gives a realistic device capable of CV quantum teleportation for non-Gaussian single photon states and even the coherent superposition of two different photon number states.

The
paper is organized as follows. In Sec. 2 we shortly describe the proposed system \cite{bar} and derive the quantum Langevin equations (QLEs). In Sec. 3, the linearization of QLEs around the semiclassical steady state is discussed and we
quantify the entanglement between the output of optical mode and the output of microwave field by using the logarithmic negativity. In Sec. 4, the fidelity of the teleportation is studied, 
while conclusions are summarized in Sec. 5.


\begin{figure}[ht]
   \centering
   \includegraphics[width=0.7\textwidth]{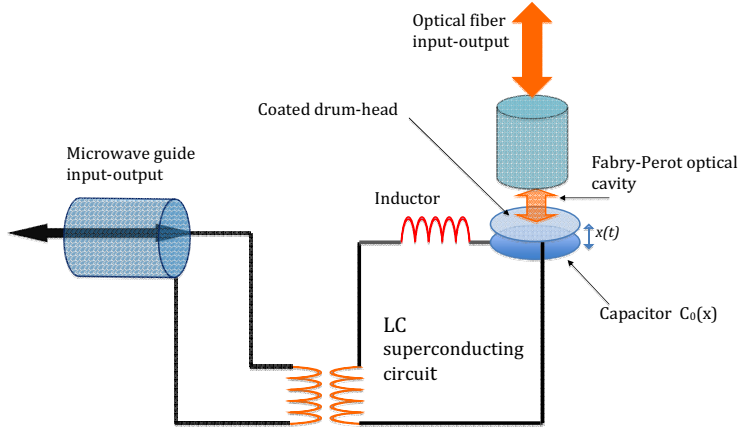} 
   \caption{ Schematic description of the device under study. A microwave transmission line source is coupled into superconducting microwave resonator. The capacitance of this resonator is modulated by a bulk mechanical resonator the motion of which modulates the frequency of an optical cavity with a fibre output coupler.   }
   \label{fig:1}
\end{figure}

\section{System dynamics}

The system studied in this work is sketched in Figure~\ref{fig:1}. We assume a mechanical
resonator (MR) which on one side is  capacitively coupled to a driven superconducting MC of resonant frequency $\omega_w$, and on the other side it is coupled to a driven OC with resonant frequency $\omega_c$. Such a system might be possible using the lumped-element superconducting resonator with free standing drum-head capacitor recently developed in Ref.\cite{Teufel}. In fact, by adding an optical coating, the drum-head capacitor could also play the role of the reflecting micro-mirror of a Fabry-Perot optical cavity formed by a second standard input mirror. The microwave and optical cavities are driven at the frequencies $\omega_{0w}=\omega_w-\Delta_{0w}$ and $\omega_{0c}=\omega_c-\Delta_{0c}$, respectively. The Hamiltonian of the coupled system reads  \cite{vitali2,regal,barzanjeh,genes}
\begin{eqnarray}\label{ham0}
H&=&\frac{\hat{p}_x^2}{2m}+\frac{m\omega^2_m \hat{x}^2}{2}+\frac{\hat{\Phi}^2}{2L}+\frac{\hat{Q}^2}{2[C+C_0(\hat{x})]}-e(t)\hat{Q}\\
&&+\hbar\omega_c a^{\dagger}a-\hbar G_{0c}a^{\dagger}a\hat{x}+i\hbar E_c(a^{\dagger}e^{-i\omega_{0c}t}-a e^{i\omega_{0c}t}),\nonumber
\end{eqnarray}
where $(\hat{x},\hat{p}_x)$ are the canonical position and momentum of a MR with frequency $\omega_m$, $(\hat{\Phi},\hat{Q})$ are the canonical coordinates for the MC, describing the flux through an equivalent indictor $L$ and the charge on an equivalent capacitor $C$, respectively; $(a,a^{\dagger})$ represent the annihilation and creation operators of the OC mode ($[a,a^{\dagger}]=1$), $E_c=\sqrt{2P_c\kappa_c/\hbar\omega_{0c}}$ is related to input driving laser, where $P_c$ is the power of the input laser and $\kappa_c$ describes the damping rate of the OC, $G_{0c}=(\omega_c/{\cal L})\sqrt{\hbar/m\omega_m}$ gives the optical radiation-pressure coupling, where $m$ is the effective mass of MR, and $\cal L$ is an effective length that depends upon the OC geometry. The coherent driving of the MC with damping rate $\kappa_w$ is given by the electric potential $e(t)=-i\sqrt{2\hbar\omega_wL}E_w(e^{i\omega_{0w}t}-e^{-i\omega_{0w}t})$, where $E_w=\sqrt{2P_w\kappa_w/\hbar\omega_{0w}}$ with $P_w$ the power of the input microwave source. We finally stress that the optical and microwave cavities might support additional degenerate modes which we ignored in writing Eq. (\ref{ham0}). This is valid as long as one assumes small cavities, in which the free spectral range (\textit{FSR}) is much larger than the mechanical frequency $\omega_ {m}$.
In this case, scattering of photons from the driven mode into other cavity modes is negligible.
This guarantees that only one cavity mode participates in the optomechanical interaction and the neighbour modes are not excited by a single central frequency input laser.

The capacitive coupling between the MC and the
MR as a function of the resonator displacement $\hat{x}$ is given by $C_0(\hat{x})$. We expand this function around the equilibrium position of the resonator corresponding to a separation
$d$ between the plates of the capacitor, with corresponding bare capacitance $C_0$,
$C_0(\hat{x})=C_0[1+\hat{x}(t)/d]$. Expanding the capacitive energy as a
Taylor series, we find to first order,
\begin{eqnarray}\label{2}
 \frac{\hat{Q}^2}{2[C+C_0(\hat{x})]}=\frac{\hat{Q}^2}{2C_{\Sigma}}-\frac{\mu}{2d C_{\Sigma}}\hat{x}(t)\hat{Q}^2,
\end{eqnarray}
where $C_\Sigma=C+C_0$ and $\mu=C_0/C_\Sigma$.
The Hamiltonian (\ref{ham0}) can be rewritten in the terms of the raising and lowering operators of the MC i.e,. $b, b^{\dagger}$($[b,b^{\dagger}]=1$) and the dimensionless position and momentum operators $\hat q=\sqrt{\frac{m\omega_m}{\hbar}}\hat x$ and $\hat p=\frac{\hat p_x}{\sqrt{\hbar m\omega_m}}$ ($[\hat x,\hat p]=i\hbar$), giving
\begin{eqnarray}\label{ham1}
H&=&\hbar \omega_w b^{\dagger}b+\hbar \omega_c a^{\dagger}a+\frac{\hbar \omega_m}{2}(\hat p^2+\hat q^2)-\frac{\hbar G_{0w}}{2}\hat q(b+b^{\dagger})^2\nonumber\\
&&-\hbar G_{0c}\hat q a^{\dagger}a-i\hbar E_w(e^{i\omega_{0w}t}-e^{-i\omega_{0w}t})(b+b^{\dagger})\nonumber\\
&&+i\hbar E_c(a^{\dagger}e^{-i\omega_{0c}t}-a e^{i\omega_{0c}t}),
\end{eqnarray}
where
\begin{eqnarray}\label{2}
b=\sqrt{\frac{\omega_w L}{2\hbar}}\hat Q+\frac{i}{\sqrt{2\hbar \omega_w L}}\hat \Phi,
\end{eqnarray}
and the coupling is given by

\begin{eqnarray}\label{2}
G_{0w}=\frac{\mu \omega_w}{2d}\sqrt{\frac{\hbar}{m\omega_m}}.
\end{eqnarray}
It is also convenient to move into interaction picture with respect to $\hbar \omega_{0w}b^{\dagger}b$ and $\hbar \omega_{0c}a^{\dagger}a$, and neglect fast oscillating terms at $\pm2\omega_{0w}, \pm 2\omega_{0c}$, so that the system's Hamiltonian becomes
\begin{eqnarray}\label{ham2}
H&=&\hbar \Delta_{0w} b^{\dagger}b+\hbar \Delta_{0c} a^{\dagger}a+\frac{\hbar \omega_m}{2}(\hat p^2+\hat q^2)-\hbar G_{0w}\hat qb^{\dagger}b\nonumber\\
&&-\hbar G_{0c}\hat q a^{\dagger}a-i\hbar E_w(b-b^{\dagger})+i\hbar E_c(a^{\dagger}-a).
\end{eqnarray}
However the dynamics of the three modes is also affected by damping and noise processes, due to the fact that each of them interacts with its own environment. We can describe them adopting a QLE treatment in which the Heisenberg equations for the system operators associated with Eq.~(\ref{ham2}) are supplemented with damping and noise terms. The resulting nonlinear QLEs are given by
\begin{eqnarray}
\dot{\hat{q}}&=&\omega_m \hat{p},\label{lan1}\\
\dot{\hat{p}}&=&-\omega_m \hat{q}-\gamma_m \hat{p}+G_{0c}a^{\dagger}a+G_{0w}b^{\dagger}b+\xi,\\
\dot{a}&=&-(\kappa_c+i\Delta_{0c})a+iG_{0c}\hat{q} a+E_c+\sqrt{2\kappa_c}a_{in},\\
\dot{b}&=&-(\kappa_w+i\Delta_{0w})b+iG_{0w}\hat{q} b+E_w+\sqrt{2\kappa_w}b_{in},\label{lan2}
\end{eqnarray}
where $\gamma_m$ is the damping rate of the mechanical mode and also $\xi(t)$ is the Brownian noise acting on the mechanical motion, with
correlation function \cite{pin2, law, gard, vitgio}

\begin{equation}\label{nois1}
\begin{array}{rcl}
\langle\xi(t)\xi(t')\rangle=\frac{\gamma_m}{\omega_m}\int \frac{d\omega}{2\pi} e^{-i\omega(t-t')}\omega\Big[\mathrm{coth}\big(\frac{\hbar \omega}{2k_B T}\big)+1\Big],
\end{array}
\end{equation}
where $k_B$ is the Boltzmann constant and $T$ is the temperature
of the reservoir of the micromechanical oscillator. As is seen from Eq.(\ref{nois1}) the Brownian noise $\xi(t)$ is not $\delta$-correlated and therefore does not describe a Markovian
process. Typically, significant optomechanical entanglement is achieved for a very high mechanical quality factor, $Q=\omega_m/\gamma_m\rightarrow \infty$. In this limit, the mechanical noise is characterized by a thermal white noise operator and $\xi(t)$ becomes a $\delta$ function \cite{vitgio} and $\langle\xi(t)\xi(t')+\xi(t')\xi(t)\rangle/2\simeq\gamma_m(2\bar{n}+1)\delta (t-t')$, where $\bar{n}=[exp(\hbar \omega_m/k_B T)-1]^{-1}$ is the mean thermal excitation number of the resonator. The optical and microwave cavity input noise operators, $a_{in}(t)$, $b_{in}(t)$,  obey the correlation functions

\begin{eqnarray}\label{coropt}
\langle a_{in}(t)a^{\dagger}_{in}\rangle&=&[N(\omega_c)+1]\delta(t-t'),\\
\langle a^{\dagger}_{in}(t)a_{in}\rangle&=&N(\omega_c)\delta(t-t'),\\
\langle b_{in}(t)b^{\dagger}_{in}\rangle&=&[N(\omega_w)+1]\delta(t-t'),\\
\langle b^{\dagger}_{in}(t)b_{in}\rangle&=&N(\omega_w)\delta(t-t'),
\end{eqnarray}
where $N(\omega_c)=[exp(\hbar \omega_c/k_B T)-1]^{-1}$ and  $N(\omega_w)=[exp(\hbar \omega_w/k_B T)-1]^{-1}$ are the equilibrium mean
thermal photon number of the optical and microwave field, respectively. One can safely assume $N(\omega_c)\approx0$ since $\hbar \omega_c/k_B T \gg 1$ at optical frequencies, while thermal microwave photons cannot be neglected in general, even at low temperatures.
\section{Linearization of QLEs}
To study the dynamics of the system we need to solve the QLEs (\ref{lan1})-(\ref{lan2}), which are  quantum nonlinear stochastic differential equations. These equations admit, however, a linearization around the semiclassical fixed points. For this purpose, one can write $a=\alpha_s+\delta a$, $b=\beta_s+\delta b$, $\hat{p}=p_s+\delta \hat p$, and $\hat{q}=q_s+\delta \hat q$.Then the fixed points are 
$$p_s=0,\,q_s=\frac{G_{0c}}{\omega_m}|\alpha_s|^2+\frac{G_{0w}}{\omega_m}|\beta_{s}|^2,$$ 
$$|\alpha_s|^2 = \frac{E_c}{\kappa_c^2+\Delta_c^2} ,\,\,\, |\beta_s|^2 = \frac{E_w}{\kappa_w^2+\Delta_w^2},$$ where $\Delta_i=\Delta_{0i}-G_{0i}q_s$ ($i=c,w$) describes the effective detuning of cavities field. We point out that, the linearization of QLEs is obtained when the radiation pressure coupling is strong, which needs very intense intracavity fields i.e., $|\alpha_s|^2>>1$ and $|\beta_s|^2>>1$, as it is shown in Ref.\cite{vitali2}. In this limit, when the stability conditions are satisfied, and choosing the phase references such that $\alpha_s$ and $\beta_s$ are real and positive, which implies that the microwave and the optical fields are phase locked, one obtains the following linear QLEs for the quantum fluctuations of the tripartite system.
 
 In terms of the OC field quantum fluctuation quadratures $\delta \hat{X}_c=(\delta a+\delta a^{\dagger})/\sqrt{2}$ and $\delta \hat{Y}_c=(\delta a-\delta a^{\dagger})/i\sqrt{2}$, the MC field quantum fluctuation quadratures $\delta \hat{X}_w=(\delta b+\delta b^{\dagger})/\sqrt{2}$ and $\delta \hat{Y}_w=(\delta b-\delta b^{\dagger})/i\sqrt{2}$, and the corresponding Hermitian input noise operators $\hat{X}^{in}_{c}=( a_{in}+ a_{in}^{\dagger})/\sqrt{2}$, $\hat{Y}^{in}_{c}=(a_{in}-a_{in}^{\dagger})/i\sqrt{2}$, $\hat{X}^{in}_{w}=( b_{in}+ b_{in}^{\dagger})/\sqrt{2}$, $ \hat{Y}^{in}_{w}=(b_{in}- b_{in}^{\dagger})/i\sqrt{2}$, the QLEs are

\begin{eqnarray}
\delta \dot{\hat{q}}&=\omega_m \delta \hat{p},\label{qles3}\\
\delta \dot{\hat{p}}&=-\omega_m \delta \hat{q}-\gamma_m \delta \hat{p}+G_{c}\delta \hat{X}_c+G_{w}\delta \hat{X}_w+\xi,\\
\delta\dot{\hat{X_c}}&=-\kappa_c \delta \hat{X}_c+\Delta_c \delta \hat{Y}_c+\sqrt{2\kappa_c}\hat{X}^{in}_c,\label{fourth}\\
\delta\dot{\hat{Y_c}}&=-\kappa_c \delta \hat{Y}_c-\Delta_c \delta \hat{X}_c+G_c \delta\hat{q}+\sqrt{2\kappa_c}\hat{Y}^{in}_c,\\
\delta \dot{\hat{X_w}}&=-\kappa_w \delta \hat{X}_w+\Delta_w \delta \hat{Y}_w+\sqrt{2\kappa_w}\hat{X}^{in}_w,\\
\delta \dot{\hat{Y_w}}&=-\kappa_w \delta \hat{Y}_w-\Delta_w\delta \hat{X}_w+G_w \delta \hat{q}+\sqrt{2\kappa_w}\hat{Y}^{in}_w,\label{qles4}
\end{eqnarray}
\\
with the new couplings $G_c=\frac{2\omega_c}{{\cal L}}\sqrt{\frac{P_c\kappa_c}{m \omega_m \omega_{0c}(\kappa_c^2+\Delta_c^2)}}
$ and $G_w=\frac{\mu \omega_w}{d}\sqrt{\frac{P_w\kappa_w}{m \omega_m \omega_{0w}(\kappa_w^2+\Delta_w^2)}}$.\\

Eqs.(\ref{qles3})-(\ref{qles4}) can compactly be written
\begin{equation}\label{compact}
\dot{\bf{u}}(t)={\bf{A u} }(t)+{\bf{n}} (t)
\end{equation}
where, 
\begin{equation}
{\bf{u}}(t)=[\delta q(t),\delta p(t),\delta X_c(t),\delta Y_c(t),\delta X_w(t),\delta Y_w(t)]^T
\end{equation}
and 
\begin{equation}
{\bf{n}}(t)=[0,\xi(t),\sqrt{2\kappa_c} X^{in}_c,\sqrt{2\kappa_c} Y^{in}_c,\sqrt{2\kappa_w} X^{in}_w,\sqrt{2\kappa_w} Y^{in}_w]^T,
\end{equation}
with the drift matrix defined by
\begin{equation}\label{driftA}
\begin{array}{rcl}
\bf{A} = \left( {\begin{array}{*{20}c}
   0 & {\omega _m } & 0 & 0 & 0 & 0  \\
   { - \omega _m } & { - \gamma _m } & {G_c } & 0 & {G_w } & 0  \\
   0 & 0 & { - \kappa _c } & {\Delta _c } & 0 & 0  \\
   {G_c } & 0 & { - \Delta_c } & { - \kappa _c } & 0 & 0  \\
   0 & 0 & 0 & 0 & { - \kappa _w } & {\Delta _w }  \\
   {G_w } & 0 & 0 & 0 & { - \Delta _w } & { - \kappa _w }  \\
\end{array}} \right).
\end{array}
\end{equation}
We can now proceed as in our previous paper Ref. \cite{bar}  bearing in mind that  the role of the optical mode and microwave mode in that paper are swapped here because the teleportation protocol we are describing is reversed with respect to Ref.\cite{bar}. In this paper, the sender is on the microwave side of the device while the receiver is in the optical side. Keeping this in mind we must exchange the optical with the microwave mode and, choosing the same detuning values  as in Ref.\cite{bar},  \textit{i.e.} 
$\Delta_c = -\Delta_w = \omega_m$.
On the other hand the intracavity optical field fluctuations (microwave field)  $\delta a(t)$ ($\delta b(t)$) and its output are related by the
usual input-output relation \cite{gard} which is characterized by $\delta a^{out}(t)=\sqrt{2\kappa_c}\delta a(t)-a_{in}(t)$ ($\delta b^{out}(t)=\sqrt{2\kappa_w}\delta b(t)-b_{in}(t)$).The output optical field $\delta a^{out}(t)$ satisfy the same commutation relation as the input optical field $a^{in}(t)$, i.e., the only nonzero commutator is $[\delta a^{out}(t),\delta a^{out}(t)^{\dagger}]=\delta(t-t')$ as well as for microwave operators $\delta b^{out}(t)$. From the continuous output field
$\delta a^{out}(t)$ ($\delta b^{out}(t)$) one can extract many independent optical modes (microwave modes), by
selecting different time intervals or, equivalently, different
frequency intervals \cite{van} depending on the details of the measurements made on the output. One can define a generic
set of \textit{N} output modes by means of the corresponding annihilation operators
\begin{equation}\label{kernel}
\begin{array}{rcl}
\delta a_k^{out}(t)=\int_{-\infty}^{t}ds g_k(t-s)\delta a^{out}(s),\,\,\,k=1,2,..,N,
\end{array}
\end{equation}
where $g_k(t)$ is the causal filter function defining the \textit{k}'th output mode. In our case the two output modes originate from two different cavities  and consequently describe two independent modes. Therefore, we can assume the following filter functions in term of the Heaviside step function $\theta(t)$ as
\begin{equation}\label{gfilter}
\begin{array}{rcl}
g_j(t)=\sqrt{\frac{2}{\tau_j}}\theta(t)e^{-(1/\tau_j+i\Omega_j) t}\,\,(j=c,w),
\end{array}
\end{equation}
where the filter functions are characterized by bandwidths $1/\tau_j$ and central frequencies, $\Omega_j$.

The entanglement between the optical-microwave output modes
is fully determined by
\begin{equation}\label{cor1}
\begin{array}{rcl}
V^{out}_{ij}(t)=\frac{1}{2}<u^{out}_i(t)u^{out}_j(t)+u^{out}_j(t)u^{out}_i(t)>.
\end{array}
\end{equation}
because Eqs. (\ref{qles3})-(\ref{qles4}) are linear and the noise is Gaussian, the variance matrix completely describes all  output moments of this system.

By using the output cavity modes Eq.(\ref{kernel}) one can derive the following general expression for
the stationary output correlation matrix \cite{genes}
\begin{equation}\label{vmat}
{\bf{V}}^{out}=\int d\omega\tilde{\bf{T}}(\omega)\Big(\tilde{\bf{M}}^{ext}(\omega)+{\bf{P}}_{out}\Big)\times{ \bf{D}}_{ext}\Big(\tilde{{\bf{M}}}^{ext}(\omega)^{\dagger}+\bf{P}_{out}\Big)\tilde{\bf{T}}^{\dagger}(\omega),
\end{equation}
where $\tilde{\bf{T}}(\omega)$ is the Fourier transforms of
\begin{eqnarray}
\bf{T}(t) = \left( {\begin{array}{*{20}c}
   {\delta (t)} & 0 & 0 & 0 & 0 & 0  \\
   0 & {\delta (t)} & 0 & 0 & 0 & 0  \\
   0 & 0 & R_c & -I_c & 0 & 0  \\
   0 & 0 & I_c & R_c & 0 & 0  \\
   0 & 0 & 0 & 0 & R_w & -I_w  \\
   0 & 0 & 0 & 0 & I_w & R_w  \\
\end{array}} \right),
\end{eqnarray}
and $\tilde{\bf{M}}^{ext}(\omega)=(i\omega\bf{I}+\bf{A})^{-1}$ with $\bf{I}$ the identity matrix, ${\bf{P}}_{out}=\mathrm{Diag}[0,0,1/2k_c,1/2k_c,1/2k_w,1/2k_w]$, the drift matrix $\bf{A}$ is given by Eq.(\ref{driftA}), ${\bf{D}}_{ext}=\mathrm{Diag}[0,\gamma_m(2\bar{n}_b+1),2\kappa_c,2\kappa_c,2\kappa_w(2N(\omega_w)+1),2\kappa_w(2N(\omega_w)+1)]$ is the diffusion matrix due to existence of noise terms in the linearized QLEs (\ref{qles3})-(\ref{qles4}), $R_j=\sqrt{2\kappa_j}Re[g_j(t)]$, $I_j=\sqrt{2\kappa_j}Im[g_j(t)]\,\,(j=c,w)$.

In order to establish the conditions under which the output of optical and microwave modes are entangled, we
consider the logarithmic negativity $E_N$, which can be defined as \cite{eis}
\begin{equation}\label{loga}
\begin{array}{rcl}
E_N=Max[0,-\mathrm{ln}(2\eta^{-})],
\end{array}
\end{equation}
where $\eta^{-}\equiv2^{-1/2}\Big(\Sigma({\bf{V}}')-\sqrt{\Sigma({\bf{V}}')^2-4 \mathrm{det} {\bf{V}}'}\Big)^{\frac{1}{2}}$ and we have used the $2\times2$ block form of the reduced CM Eq.(\ref{vmat}) as
\begin{equation}\label{loga}
\begin{array}{rcl}
{\bf{V}}'=\left(
     \begin{array}{cc}
       \bf{B} & \bf{C} \\
      { \bf{C}}^T & {\bf{B}}' \\
     \end{array}
   \right),
\end{array}
\end{equation}
where
\begin{eqnarray}
 \Sigma(\bf{V}')\equiv \mathrm{det} \bf{B}+\mathrm{det} \bf{B}'-2\mathrm{det} \bf{C},
\end{eqnarray}
and
\begin{eqnarray}
&&\bf{B}=\left(
           \begin{array}{cc}
             V_{33} & V_{34} \\
             V_{34} & V_{44} \\
           \end{array}
         \right)
,
\bf{B}'=\left(
           \begin{array}{cc}
             V_{55} & V_{56} \\
             V_{56} & V_{66} \\
           \end{array}
         \right), \bf{C}=\left(
           \begin{array}{cc}
             V_{35} & V_{36} \\
             V_{45} & V_{46} \\
           \end{array}
         \right).
\end{eqnarray}
\begin{figure}[ht]
   \centering
   \includegraphics[width=.57\textwidth]{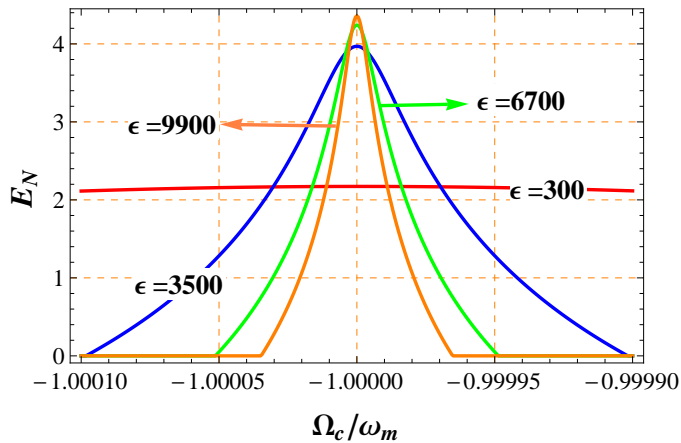} 
   \caption{\small{$E_{N}$ at four different values of the normalized inverse bandwidth $\epsilon=\tau\omega_m$ vs. the normalized frequency $\Omega_c/\omega_{m}$, at fixed central frequency of the microwave output mode $\Omega_w=\omega_m$. $\lambda_{0c}=810$ nm and power $P_c=3.4$ mW, with $\omega_{0w}/2\pi=10$ GHz and microwave input power $P_w=42$ mW.}
   }
   \label{fig:2}
\end{figure}
To determine the best entanglement between the output of optical-microwave modes, we have plotted the logarithmic negativity versus the normalized central frequency $\Omega_c/\omega_m$ at  four different values of the normalized inverse bandwidth $\epsilon= \epsilon_w=\epsilon_c=\tau \omega_m$ at $\Delta_w=\omega_m$, $\Delta_c=-\omega_m$ and $\Omega_w=\omega_m$ in Figure~\ref{fig:2}, where we have assumed an experimental situation representing a feasible extension of the scheme of Ref.~\cite{Teufel}, i.e., we have assumed a lumped-element superconducting circuit with a free standing drum-head capacitor, which is then optically coated to form a micromirror of an additional optical Fabry-Perot cavity. We have taken parameters similar to that of Ref.~\cite{Teufel} for the MC and MR, that is a MR with $\omega_m/2\pi=10$ MHz, $Q=15\times 10^4$, and a MC with $\omega_w/2\pi=10$ GHz, $\kappa_w=0.04\omega_m$, driven by a microwave source with power $P_w=42$ mW. The coupling between the two is determined by the parameters $d=100$ nm, $\mu=0.013$. We have considered a lower mechanical quality factor, and resonator higher mass $m = 10$ ng than that of Ref.~\cite{Teufel}, in order to take into account the presence of the coating, which typically worsens the mechanical properties. We have then assumed an OC of length ${\cal L} = 1$ mm and damping rate $\kappa_c=0.04 \omega_m$ driven by a laser with wavelength $\lambda_{0c}=810$ nm and power $P_c=3.4mW$. As it is shown in Figure~\ref{fig:2} the higher $\epsilon$ (small bandwidths)  the larger the stationary entanglement appears around the blue detuned sideband at $\Omega_c= -\omega_m$.
Thus, It could be possible to control the entanglement of the microwave-optical modes by varying the detection bandwidth $\tau^{-1}$. From the experimental point of view this means that one can obtain an effective entanglement distillation by appropriately filtering the output fields. Similar results have also been obtained in the case of entanglement of the output of optical modes and the movable mirror \cite{genes}.

\section{CV teleportation between the optical-microwave output modes}
\subsection{Single photon state}
We have seen that the vibrational mode of the MR realizes an effective 
entanglement between optical and microwave output modes. Since optical (microwave)
traveling wave fields (output modes) are typically used for CV quantum
information applications, this fact suggests the possibility of using the output microwave fields (optical fields) in the manipulation and storage of CV quantum information. 

The first experimental demonstration of a CV quantum information
protocol was the quantum teleportation of an unknown
coherent state of an optical mode onto another optical mode
illustrated in Ref.\cite{furusawa1}. Quantum teleportation requires the use of shared entanglement between
two distant stations (the quantum channel), Alice and Bob, and of a classical channel for the transmission of the results of the Bell measurement from Alice to Bob. The coupling between the microwave and optical output modes establishes the required quantum channel, i.e.,
the shared entangled state between the output of optical mode in Alice's hand and that of microwave mode at Bob's station. 
 The teleportation scheme in the current setting is the same as that the proposed in Ref.\cite{bra} but exchanging the role of Alice and Bob. In the scheme proposed in this paper  a single photon number state  of the radiation
field in the microwave cavity is prepared by a verifier (Victor) in a source cavity and then emitted towards Bob's station. To implement 
a teleportation protocol Bob needs to perform a joint measurement on the output of the source cavity and his part of the entangled microwave-optical modes.  

He can do this by mixing the output of the source cavity and his entangled beam on balanced beam splitter with the output of the microwave mode. The two outputs of the beam splitter are then subject to a homodyne measurement, using two IQ mixers,  with a pulsed local oscillator mode matched to the single photon source cavity. The integrated homodyne current then produces two measurement results, $ X_+$ and $ P_-$.   Currently the quantum efficiency of the homodyne measurement is not high due to the need to amplify the signal prior to using the IQ mixer however the use of phase dependent amplifiers, such as Josephson parametric amplifiers, should improve the quantum efficiency \cite{Lehnert_JJA}.

These results are then passed through a classical channel to Alice who completes the protocol by implementing conditional displacements of her component of the shared entangled beams. These displacements will need to be done using a pulsed local oscillator synchronous with and phase locked to that used for the microwave measurements at Bob's station.   Upon receiving this information, Alice displaces her part of entangled state (the output of optical mode) as follows:
$\hat X_{c}^{out}\rightarrow\hat X_{c}^{out}+\sqrt{2}X_+$ and $\hat P_{c}^{out}\rightarrow\hat P_{c}^{out}-\sqrt{2}P_-$. We emphasize
that Alice and Bob do not assume any
prior knowledge of the input state and adhere to
unity-gain teleportation, so that the teleporter
does not have any restriction regarding the specific
family of quantum states it can faithfully teleport.

\begin{figure}[ht]
   \centering
   \includegraphics[width=.57\textwidth]{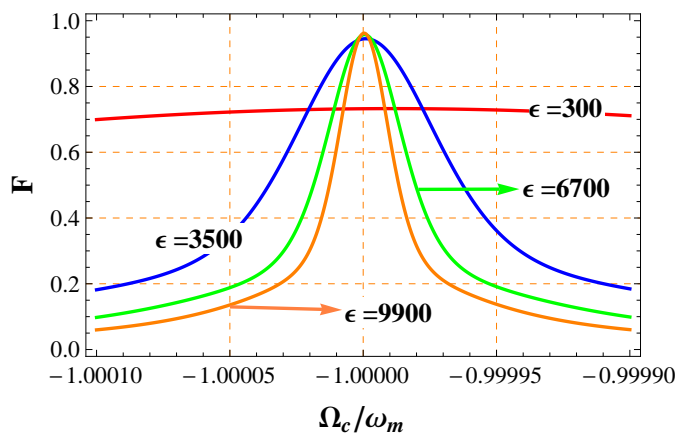} 
   \caption{Teleportation fidelity $F$ at four different values of $\epsilon=\tau\omega_m$ vs $\Omega_c/\omega_m$ and for single photon state $|1\rangle$. The other parameters are as in Fig.~\protect\ref{fig:2}.
   }
   \label{fig:3}
\end{figure}

To quantify the quality of teleportation protocol in the system under study one can use the fidelity that in the case of a pure state $|\psi_{in}\rangle$, it is given by $F=\langle\psi_{in}|\rho_{out}|\psi_{in}\rangle$, where $\rho_{out}$ is the output state of the protocol. In our case, the non-Gaussian single photon state can only be teleported and retrieved at the output port when $F>F_{th}$ \cite{ban}, a threshold bound $F_{th}=2/3$ known as the no-cloning limit. Thus we have a practical criterion to determine the successful transfer of single photon number state.

 We thus restrict the discussion to the case when the input state
is a single photon number state $|1\rangle$ with characteristic function
\begin{eqnarray}\label{phi}
\phi_{n=1}^{in}(\lambda)&=&L_{n=1}(|\lambda|^2)\mathrm{exp}(-|\lambda|^2/2)=(1-|\lambda|^2)\mathrm{exp}(-|\lambda|^2/2),
\end{eqnarray}
were $L_n(x)$ is the Laguerre polinomial of degree $n$. Our quantum channel is a Gaussian channel with corresponding characteristic function $\Phi^{ch}(\vec\xi)=\mathrm{exp}(-\vec\xi^T {{\bf V}}^{ch}\vec\xi/4+i\vec d^T\vec\xi)$ (where $\vec\xi^T=(X^{out}_c, Y^{out}_c, X^{out}_w, Y^{out}_w) $ is the vector in the phase-space of variables and ${{\bf V}}^{ch}$ is the reduced covariance matrix ${{\bf V}}^{'}$). We also  assume that Alice and Bob share a zero-displacement state, implying $\vec d=\vec 0$. Thus the fidelity of the teleportation can be written in terms of the channel and the input state as \cite{pir}
\begin{eqnarray}\label{f1}
F=\pi^{-1}\int d^2\eta|\phi^{in}(\eta)|^2[\phi^{ch}(\eta^*,\eta)]^*.
\end{eqnarray}
 By plugging Eq.(\ref{phi}) into Eq.(\ref{f1}) and after some algebraic rearrangement we obtain
\begin{eqnarray}\label{fidchar}
F&=&\pi^{-1}\int d^2\eta \Big(1-|\eta|^2\Big)^2\mathrm{exp}(-\vec\mu^T {{\bf \Gamma}}\vec\mu)
\end{eqnarray}
where $ \vec\mu^T=[\eta_I,-\eta_R] $, $(\eta=\eta_R+i\eta_I)$   and
${{\bf \Gamma}}=2\bf{V}_{\rm coh}+\bf{ZBZ}+\bf{ZC}+\bf{C}^T\bf{Z}+\bf{B}',$
$\bf{Z}=\left(\begin{array}{cc}1 & 0 \\0 & -1 \\\end{array}\right)$
$,\,\, {\bf{V}}_{\rm coh}=\frac{1}{2}\left(
     \begin{array}{cc}
       1 & 0 \\
       0 & 1 \\
     \end{array}
   \right).$

 The fidelity of teleportation after performing the integral is given by
\begin{eqnarray}
\textit{F}=\frac{1}{\sqrt{\mathrm{det}\bf{\Gamma}}}\Big(1+\frac{1}{\mathrm{det}\bf{\Gamma}}[\frac{1}{2}-(\Gamma_{11}+\Gamma_{22})]+\frac{3[\Gamma_{11}^2+\Gamma_{22}^2+2\Gamma_{12}^2]}{4(\mathrm{det}\bf{\Gamma})^2}\Big).
\end{eqnarray}

Figure~\ref{fig:3} shows the fidelity of the teleportation protocol between the optical-microwave output modes versus normalized central frequency $\Omega_c/\omega_m$ in the case of four different values of $\epsilon$ with the same data of Fig.~\ref{fig:2}. 
Clearly, at $\Delta_c=\omega_m$, $\Delta_w=-\omega_m$ and $\Omega_w=\omega_m$ the fidelity is highly
peaked around $\Omega_c=-\omega_m$ where is larger than the no-cloning limit. Furthermore, similar to the logarithmic negativity, the fidelity of the protocol can be controlled by varying the frequency bandwidth $\tau^{-1}$. It is interesting that for large enough values of frequency bandwidth $\tau^{-1}$ the teleportation
fidelity is always greater than the no-cloning
limit of 2/3 very near $\Omega_c=-\omega_m$, which shows the great practical potential of this system to teleport a single photon state.

We pointed out that in the current scheme Alice receives the optical output mode and Bob gets an output microwave mode, by considering the symmetry of the system  is apparent that exchanging the microwave and optical modes one is able to propose the inverse protocol, i.e. Alice receives form Victor the single optical photon and transfer it to Bob who gets a  single photon state at microwave frequency centered at $\Omega_w=\omega_m$.
\subsection{Superposition state}
\begin{figure}[ht]
   \centering
   \includegraphics[width=.57\textwidth]{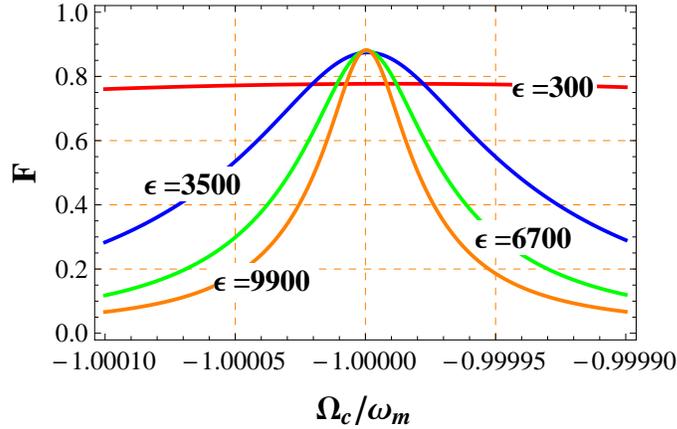} 
   \caption{Teleportation fidelity $F$ at four different values of $\epsilon=\tau\omega_m$ vs $\Omega_c/\omega_m$ for the coherent superposition of number states $\frac{1}{\sqrt{2}}(|0\rangle+|1\rangle)$. The other parameters are as in Fig.~\protect\ref{fig:2}.
   }
   \label{fig:4}
\end{figure}

 Let us now consider the special case in which the input state
is a coherent superposition of number states $|\psi^{in}\rangle=\frac{1}{\sqrt{2}}(|0\rangle+|1\rangle) $. The Wigner characteristic function of this state is given by 
\begin{equation}
\phi^{in}(\eta)=Tr(\rho e^{\eta a^{\dagger}-\eta^* a})=\frac{1}{2}(2-|\eta|^2+2\eta_R)e^{-|\eta|^2/2}.
\end{equation} 
The fidelity of teleportation can be obtained by substituting $\phi^{in}$ into Eq.(\ref{f1}) which reads as
\begin{eqnarray}
F=\frac{1}{4\pi}\int d^2\eta \Big(4+|\eta|^4+4\eta_R^2-&&4|\eta|^2+8\eta_R-4|\eta|^2\eta_R)
\mathrm{exp}(-\vec\mu^T {{\bf \Gamma}}\vec\mu)
\end{eqnarray}
Performing the integral we obtain the fidelity of teleportation as follows 
\begin{eqnarray}\label{ff22}
\textit{F}=\frac{1}{4\sqrt{\mathrm{det}\bf{\Gamma}}}\Big(4+\frac{1}{\mathrm{det}\bf{\Gamma}}[\frac{1}{2}-2\Gamma_{22}]+\frac{3[\Gamma_{11}^2+\Gamma_{22}^2+2\Gamma_{12}^2]}{4(\mathrm{det}\bf{\Gamma})^2}\Big)\nonumber.\\
\end{eqnarray}
The teleportation fidelity for the superposition states Eq.(\ref{ff22}) is shown in Figure~\ref{fig:4}. 
\section{Conclusion}
We have proposed a scheme for the realization of the CV teleportation of a single photon state and the superposition of two number states of radiation between the outputs of
optical and microwave modes by means of a micro mechanical resonator. As we have shown, the mechanical resonator leads to the entanglement between an output of optical mode and an output of microwave mode. This entanglement can be used as a realistic Gaussian quantum channel to approach the CV quantum teleportation. We have shown that for experimentally feasible parameters and at optical and microwave frequencies the protocol is identical to the standard Braunstein-Kimble protocol \cite{bra}, and the proposed scheme is able to teleport non-Gaussian number states and its superpositions with fidelity well above the no-cloning limit. 

From experimental point of view the realization of this teleportation experiment resides on the possibility of performing a homodyne measurement at microwave frequency. Assuming a sufficiently good efficiency for the detectors, this can be obtained considering a microwave beam splitter as the one in Ref. \cite{nguyen}, where it was used to show the interference between two single microwave photons at different frequencies; and the single microwave photon counter to perform the homodyne detection could be the one introduced in Ref.\cite{mcdermott} to measure the coincidence counting statistics of microwave photons. The teleported result could be detected by a tomographic apparatus as the one in Ref. \cite{mlynek} used to reconstruct the single photon Fock state at optical wavelength.

It is worth mentioning here that the same result could be obtained if, instead of a microwave cavity, the second cavity were another optical cavity at different frequency. In that case the device would be able to convert single photons at different frequencies at will.

\end{document}